\documentstyle[12pt]{article}
\newcommand{\sect}[1]{\setcounter{equation}{0}\section{#1}\indent}

\begin{document}
\def\bbox{{\,\lower0.9pt\vbox{\hrule \hbox{\vrule height 0.2 cm
\hskip 0.2 cm \vrule height 0.2 cm}\hrule}\,}}
%%%%%%%%%%%%%%%%%%%%%%%%%%%%%%%%%%%%%%%%%%%%%%%%
\def\a{\alpha}
\def\b{\beta}
\def\g{\gamma}
\def\G{\Gamma}
\def\d{\delta}
\def\dag{\dagger}
\def\D{\Delta}
\def\e{\epsilon}
\def\h{\hbar}
\def\ve{\varepsilon}
\def\z{\zeta}
\def\t{\theta}
\def\vt{\vartheta}
\def\r{\rho}
\def\vr{\varrho}
\def\k{\kappa}
\def\l{\lambda}
\def\L{\Lambda}
\def\m{\mu}
\def\n{\nu}
\def\o{\omega}
\def\O{\Omega}
\def\s{\sigma}
\def\vs{\varsigma}
\def\S{\Sigma}
\def\vphi{\varphi}
\def\av#1{\langle#1\rangle}
\def\pa{\partial}
\def\p{\parallel}
\def\na{\nabla}
\def\hg{\hat g}
\def\un{\underline}
\def\ov{\overline}
\def\cF{{\cal F}}
\def\cG{{\cal G}}
\def\cN{{\cal N}}
\def\Hsl{H \hskip-8pt /}
\def\Fsl{F \hskip-6pt /}
\def\cFsl{\cF \hskip-5pt /}
\def\ksl{k \hskip-6pt /}
\def\pasl{\pa \hskip-6pt /}
\def\tr{{\rm tr}}
\def\tcF{{\tilde{{\cal F}_2}}}
\def\tg{{\tilde g}}
\def\ta{{\tilde\alpha}}
\def\tp{{\tilde\psi}}
\def\shalf{\frac{1}{2}}
\def\nn{\nonumber \\}
\def\w{\wedge}
\def\ra{\rightarrow}
\def\la{\leftarrow}
\def\be{\begin{equation}}
\def\ee{\end{equation}}
\newcommand{\brr}{\begin{eqnarray}}
\newcommand{\err}{\end{eqnarray}}

\def\cmp#1{{\it Comm. Math. Phys.} {\bf #1}}
\def\cqg#1{{\it Class. Quantum Grav.} {\bf #1}}
\def\pl#1{{\it Phys. Lett.} {\bf B#1}}
\def\prl#1{{\it Phys. Rev. Lett.} {\bf #1}}
\def\prd#1{{\it Phys. Rev.} {\bf D#1}}
\def\prr#1{{\it Phys. Rev.} {\bf #1}}
\def\prb#1{{\it Phys. Rev.} {\bf B#1}}
\def\np#1{{\it Nucl. Phys.} {\bf B#1}}
\def\ncim#1{{\it Nuovo Cimento} {\bf #1}}
\def\jmp#1{{\it J. Math. Phys.} {\bf #1}}
\def\aam#1{{\it Adv. Appl. Math.} {\bf #1}}
\def\mpl#1{{\it Mod. Phys. Lett.} {\bf A#1}}
\def\ijmp#1{{\it Int. J. Mod. Phys.} {\bf A#1}}
\def\prep#1{{\it Phys. Rep.} {\bf #1C}}

%%%%%%%%%%%%%%%%%%%%%%%%%%%%%

\begin{titlepage}
\setcounter{page}{0}

\begin{flushright}
COLO-HEP-453 \\
hep-th/0101200 \\
January 2001
\end{flushright}

\vspace{5 mm}
\begin{center}
{\large Boundary String Field Theory, the Boundary State Formalism and D-Brane Tension}
\vspace{10 mm}

{\large S. P. de Alwis\footnote{e-mail: 
dealwis@pizero.colorado.edu}}\\
{\em Department of Physics, Box 390,
University of Colorado, Boulder, CO 80309.}\\
\vspace{5 mm}
\end{center}
\vspace{10 mm}

\centerline{{\bf{Abstract}}}
Recently a boundary string field theory  that had been proposed some time
ago, was used to calculate correctly the ratios of D-brane (both BPS and 
non-BPS)  tensions. We discuss how this work is related to the boundary state formalism and open string closed string duality, and argue that the latter clarifies why the correct tension ratios are obtained in these recent calculations. 

\end{titlepage}
\newpage
\renewcommand{\thefootnote}{\arabic{footnote}}
\setcounter{footnote}{0}

\setcounter{equation}{0}            
\sect{Introduction}
The notion that the effective low energy field theory of string modes is
given essentially by the partition function of the corresponding string sigma model,
has a long history \cite{ft, ew}. In particular Witten \cite{ew} suggested that (a certain expression derived from) the partition function for the bosonic open string on a disc, with a boundary  term giving the contribution of open string fields,
 was a candidate for open string field theory. This has been called boundary string field theory (BSFT) in the recent literature. This theory has been used to study Sen's arguments\footnote{For a review see \cite{sen}.} on tachyon condensation \cite{shat, kmm1}.
 The most remarkable aspect of this  is that
with the quadratic profile for the tachyon, for which one could calculate
the partition function exactly \cite{ew},  exact agreement with the expected
ratio of bosonic D-brane tensions \cite{kmm1} was obtained. In the superstring case,
where it was postulated that the  
 BSFT was given exactly by the partition function, one also obtained \cite{kmm2} the correct tension ratio \cite{sen} between the BPS and non-BPS branes.

Now in these calculations the emergence of the right ratio seems to be
somewhat mysterious, arising from the constant term in the asymptotic
behavior of the Gamma function! The purpose of this note is to relate this
calculation to previous calculations of the D-brane tension; in particular
to the boundary state formalism and the t-duality argument. Many of the ingredients for our discussion have been presented 
in papers by other authors. In particular the use of open/closed string duality in the compactified string formalism to calculate the normalization of the boundary state in the bosonic case has been done in \cite{frau, ers} (see also \cite{hkms}). The relation of the boundary tachyon coupling RG flow to
the change in boundary conditions from Neumann to Dirichlet has been discussed in \cite{hkm}. What we do here is to  put all these ideas together to  demonstrate that  open/closed string duality leads to the correct D-tension formulae (for BPS and non-BPS branes) and to elucidate the relation to T-duality and the BSFT method. 

\sect{Boundary state formalism, open/closed string duality }

The boundary state formalism was used to discuss open closed string duality in a series of papers by the authors of \cite{clny} and in \cite{pc}. We will follow closely the development given in  the former reference and
that in \cite{bg} where the boundary state formalism was first applied to
non-BPS branes.

Let us first quote some standard formulae in order to establish our conventions.
The string action (in flat space) is given by (setting $\a '=2$)
\be\label{string} S={1\over 4\pi}\int dwd\ov w\left (\pa_w X^{\mu}\pa_{\ov w}X_{\mu}+\psi^{\mu}\pa_{\ov w}\psi_{\mu}+\tilde\psi^{\mu}\pa_{ w}\tilde\psi_{\mu}\right ).\ee

We will work in the double Wick rotated formalism as in \cite{bg} in order
to avoid introducing the ghost sector. So we will take $\mu = 0,1,...7$ to be directions transverse to the light cone directions which are taken to be
$\mu =8,9$. (i.e. we have Wick rotated $X^9\ra iX^9, X^0\ra -iX^0)$. The disadvantage is that we are really now discussing D-instantons rather than
D-branes and we are confined to $p\le 7$ but the results obviously will apply to D-branes after Wick rotating back. The procedure should be completely equivalent to working with ghosts and is simply a much less cumbersome way of getting at the right results which in fact should actually be valid for all $p\le 9$. 
We will also put the system on a torus with radii $R^{\mu}$ since one of our objects is to  demonstrate the relation to T-duality.

\un{Closed string channel}:\\

The closed string solution is
\be X^{\mu} = x_0^{\mu}-2i{n^{\mu}\over R^{\mu}}\tau -w^{\mu}R^{\mu}\s +i\sum_{m\ne 0}{1\over m}\left (\a_m^{\mu}e^{im(\s+i\tau )}+\tilde\a_m^{\mu}e^{-im(\s-i\tau )}\right )\ee

\brr \psi(w) &=&\sum_{r\e {\cal Z}+\nu}\psi_r^{\mu}e^{ir(\s+i\tau )}\nn
\tilde\psi (\ov w)&=&\sum_{r\e {\cal Z}+\nu}\tilde\psi_r^{\mu}e^{-ir(\s-i\tau )}\err
The (anti) commutation relations are
\brr & &[\a_m^{\mu},\a_n^{\nu}]=[\tilde\a_m^{\mu},\tilde\a_n^{\nu}]=
m\d^{\mu\nu}\d_{m,-n}, ~~[\a_m^{\mu},\tilde\a_n^{\nu}]=0,\nn & &\{\psi_r^{\mu},\psi_s^{\nu}\}=\{\tilde\psi_r^{\mu},
\tilde\psi_s^{\nu}\}=\d_{r,-s}\d^{\mu\nu},~~\{\psi_r^{\mu},\tilde\psi_s^{\nu}\}=0\err
The closed string Hamiltonian is
\brr H=L_0+\ov L_o &=&\sum{n^2\over R^2}+{1\over4}\sum{w^2R^2}+\sum_{n=1}^{\infty}(\a_m^{\mu}\a_{m\mu}+\tilde\a_m^{\mu}\tilde\a_{m\mu})\nn &+&\shalf\sum_{r={\cal Z}+\nu}(r:\psi^{\mu}_{-r}\psi_{r\mu}:+r:\tilde\psi^{\mu}_{-r}\tilde\psi_{r\mu}:)+C_c\err

with $C_c=-1$ in the NSNS sector and $C_c=0$ in the RR sector.

In the closed string channel we need to calculate the amplitude
\be\label{closedamp}Z^c(l)=<Dp|e^{-2\pi lH_c}|Dp>\ee
for the emission and absorption of a closed string state between two (or the same) $Dp$ branes. The latter are constructed in terms of boundary states
$|Bp,\eta>$ that satisfy the following boundary conditions \cite{clny, pc}.
\brr\label{bconditions} & &(\a_m+\ta_m)^{\mu}|Bp,\eta>=0,~(\psi_r+i\eta\tp_{-r})^{\mu}|Bp,\eta>=0,~~\mu =0,...,p\nn
& &(\a_m-\ta_m)^{\mu}|Bp,\eta>=0,~(\psi_r-i\eta\tp_{-r})^{\mu}|Bp,\eta>=0,~~\mu =p+1,...,7\err
where $\eta =\pm$ for the two spin structures.
The solution to these conditions is
\begin{equation}\label{nsb}
|Bp,\eta >_i=g_p^i\exp\left\{\sum_{n=1}^{\infty}
{1\over n}\a_{-n}^{\mu} T_{\mu\nu}\tilde\a_{-n}^{\nu}+i\eta\sum_{r={\cal Z}^+-\shalf}\psi_{-r}^{\mu}T_{\mu\nu}\tilde\psi^{\nu}_{-r}\right\}|Bp,\eta >_i^0
\end{equation}
Here the index $i$ goes over NSNS and RR sectors and $T$ is a diagonal
$8\times 8$ matrix with -1 for the Neumann ($\mu =0,...,p$) directions and +1 for the Dirichlet ($\mu =p+1,...,7$) directions. The ground state is an eigenstate of momentum with eigenvalue zero in the  N directions and is an eigenstate of position in the D directions. Also we have \cite{clny}
$${1\over 4i}g_p^{RR}=g_p^{NSNS}\equiv g_p.$$

Using these formulae we can calculate the amplitude $Z$ in the various sectors and we get
\be\label{closedz}<Bp\pm |e^{-2\pi lH_c}|Bp\pm >_{NSNS}=g_p^2f(R){\t_{00}(0,2il)^4\over\eta (2il)^{12}}\ee

\be{}<Bp\pm |e^{-2\pi lH_c}|Bp\mp >_{NSNS}=g_p^2f(R){\t_{01}(0,2il)^4\over\eta (2il)^{12}}\ee
\be{}<Bp\pm |e^{-2\pi lH_c}|Bp\pm >_{RR}=-16g_p^2f(R){\t_{10}(0,2il)^4\over\eta (2il)^{12}}\ee
\be{}<Bp\pm |e^{-2\pi lH_c}|Bp\mp >_{RR}=-16g_p^2f(R){\t_{11}(0,2il)^4\over\eta (2il)^{12}}=0\ee
where
\be\label{}f(R)\equiv\prod_{\mu\e\p}\t_{oo}(0,il{(R^{\mu})^2\over 2})\prod_{\mu\e\perp}\t_{oo}(0,il{2\over (R^{\mu})^2}).\ee
In the above we have put ${\p}=\{\mu=0,...,p\},~~{\perp}=\{\mu=p+1,...,9\}$.

\un{Open string channel}\\
The solution with N boundary conditions  is (omitting the space time index)
\be X=x+4i{n\over R}\tau+i2\sum_{m\ne 0}{\cos n\s\over m} e^{-m\tau}\a_m,\ee
where $m$ is an integer characterizing the Kaluza-Klein momentum while that for D boundary conditions is
\be X=x+2wR+i2\sum_{m\ne 0}{\sin n\s\over m} e^{-m\tau}\a_m\ee
where $w$ is the winding number.
Note that for simplicity we are taking the two branes to be coincident.
The Hamiltonian is 
\be H=L_o=2\sum_{\p}{n^2\over R^2}+\shalf\sum_{\perp}(wR)^2+\sum_{m=1}^{\infty}\a_{-m}^{\mu}\a_{m\mu}+\shalf\sum_{r={\cal Z}+\nu}r:\psi^{\mu}_{-r}\psi_{r\mu}:+C_o\ee
where $C_o=-\shalf /0$ in NS/R sectors.

Calculating  now in the open string channel we have
\be Z^o_{\a\b}=\tr_{\a}e^{-2\pi tH_o}e^{i\pi\b F}=\prod_{\mu\e\p}\t_{oo}(0,{i4t\over (R^{\mu})^2})\prod_{\mu\e\perp}\t_{oo}(0,it(R^{\mu})^2){\t_{\a\b}(0,it)^4\over \eta (it )^{12}}\ee
where $t={1\over 2l}$.
Using the modular transformation properties of the theta functions we get
\be\label{zopentrans} Z_{\a\b}={\prod_{\p}R^{\mu}/\sqrt2\over
\prod_{\perp}R^{\mu}/\sqrt 2 }{1\over 2^4}{\sqrt l^{\#\p+\#\perp}\over\sqrt l^8}f(R){\t_{\a\b}(0,it)^4\over \eta (it )^{12}}\ee
 
Now in conformal field theory we would have just equated this to
the corresponding closed string expression (\ref{closedz}) if we  had only the eight non-light cone directions. However in string theory even in the light cone gauge the zero modes go over ten directions so that in the above $\#\p+\#\perp =10$ rather than 8. So the actual equation is 
\be\int{dt\over 2t}Z_{\a\b}^o(t)=\int dl Z_{\b\a}^c(l).\ee
This then gives\footnote{Note that since $\int{dt\over 2t}=\int {dl\over 2l}$ one gets an extra factor of ${1\over 2l}$ inside the $l$ integral that cancels the extra factor of $l$ in the numerator of (\ref{zopentrans}).} after restoring $\a '$,
\be\label{norm}g_p^2={\prod_{\perp}R^{\mu}\over\prod_{\p}R^{\mu}}{\a '^{4-p}\over 2^5}.\ee

Actually the properly projected closed string sectors requires that we take
the appropriate left and right GSO projections so that the correct boundary states are\footnote{See for example \cite{gab} for a recent review.},
\brr\label{gsostate}|Bp>_{NSNS}={1\over\sqrt 2}(|Bp+>_{NSNS}-|Bp+>_{NSNS})\err
Thus we have
\be\int dl<Bp|e^{-2i lH_c}|Bp>_{NSNS}=\int dt\tr_{NS-R}e^{-2\pi tH_o}\ee
The open string channel is not GSO projected and so there is an open string
tachyon. Thus the boundary state (\ref{gsostate})  in fact represents the
non-BPS D-brane (for odd (even) $p$ in type IIA (IIB))\footnote{For an explanation of why the other values of $p$ do not give non-BPS states see for instance \cite{gab}.}. The BPS D brane states are given by
\be\label{D} |Dp>={1\over\sqrt 2}(|Bp>_{NSNS}+|Bp>_{RR})\ee
where $|Bp>_{RR}={1\over\sqrt 2}(|Bp+>_{RR}+|Bp->_{RR})$ is the GSO projected RR state. Then we have
\be\int dl<Dp|e^{-2i lH_c}|Dp>_{NSNS}=\int dt\tr_{NS-R}\shalf (1+(-1)^F)e^{-2\pi tH_o}\ee
so that the open string tachyon is projected out. Because of the zero modes  and the GSO projection in the RR sector the construction is consistent only for $p$ even (odd)
in the IIA (IIB) theory.

\sect{D brane tension}

Let us compute the overlap of the BPS and non-BPS boundary states  with the one graviton/dilaton state (with zero momentum). (For definiteness we will take the tensor component  in say the 00 direction which is longitudinal for all $p$).
\be\label{grav} |g> =\psi_{-\shalf}\tp_{-\shalf}|0>_{NSNS}\ee
The overlap with the non-BPS state gives
\be\label{overlapB} <g|Bp>_{NSNS}=i\sqrt 2g_p\ee
while that with the BPS state is (see (\ref{D}))
\be\label{overlapD} <g|Dp> ={1\over \sqrt 2}<g|Bp>_{NSNS}=ig_p\ee
The coupling to the graviton should be proportional to the tension of the brane and so the above is just Sen's result \cite{sen2} (see also \cite{frau2}) that the non-BPS tension is
$\sqrt 2$ times the corresponding BPS tension.
Let us now derive the exact formula for the tension of the branes.
On the one hand we have the formula (\ref{norm}) for $g_p$ and on the other
we have argued above that it must be proportional to the tension. Since 
$g_p$ is dimensionless we may therefore write,
\be g_p=\left ({R_0...R_{p}\over R_{p+1}...R_9}{\a '^{4-p}\over 2^5}\right )^{\shalf}=CT_p\prod_{\mu =0}^{p}2\pi R_{\mu}\ee
So we have
\be {T_p\over T_{p-1}}{\prod_{\mu =0}^{p}2\pi R_{\mu}\over \prod_{\mu =0}^{p-1}2\pi R_{\mu}}={g_p\over g_{p-1}}={R_p\over\sqrt{\a '}}\ee
giving,
\be {T_p\over T_{p-1}}={1\over 2\pi\sqrt{\a '}}\ee

This is  exactly the formula obtained from T-duality \cite{gt, sda} and  is  to be expected since the passage from N boundary conditions to D boundary conditions can be effected by T-duality. Indeed
 the $R$ dependence in $g_p$ reflects  that as noted in \cite{ers}, since for each such switch of boundary conditions
$R\ra {\a '\over R}$ in (\ref{norm}) . The absolute
normalization can be fixed as in \cite{sda} by defining the coupling constant $g$ to be the ratio of the F string to the D string i.e. writing
$T_1=g^{-1}{1\over 2\pi\a '}$. Then we get\footnote{The original formula derived by Polchinski \cite{jp} had a factor of the gravitational coupling in it since it was derived by comparing the string calculation to the low energy effective action. The  formula in (\ref{tension}) was  derived in \cite{sda}. The comparison between the two fixes the gravitational coupling in terms of the string scale and the string coupling. The latter also follows from the Dirac quantization rule.}
 
\be\label{tension}C={g\a '^{5}\over 2\pi \sqrt{R_0...R_{9}}}~ {\rm and} ~ T_p={g^{-1}\over (2\pi )^p\sqrt{\a '}^{p+1}}.\ee

If we remain within a particular theory (say IIA) then we can start with $p=9$ (which in this case is non-BPS brane) and then change boundary conditions in one direction to get the (BPS) 8-brane etc. What the discussion of the above two paragraphs shows is that as one goes down in $p$ the value of the normalization constant (after acounting for the world volume factor) effectively changes by ${\sqrt 2}{1\over 2\pi\sqrt{\a'}}$ when we go from non-BPS to BPS D-branes.

\sect{Relation to BSFT} In this section we will show following \cite{clny}
how the boundary state can be written as a path integral. In particular we
will show that the normalization coefficient  will be given by the integral
over the modes of the sigma model field on the boundary of a disc of the classical action. This then relates the previous calculation to that of 
\cite{kmm2} (see also \cite{kl, ttu}). 

The idea is to first construct the boundary state corresponding to having
N boundary conditions in all directions. The boundary state with some D directions is then going to be obtained by adding a `tachyon' term that will result in RG flow to a new fixed point that will correspond to D boundary conditions as in \cite{hkm}. 

Let us first just consider the bosonic sector and focus on one coordinate. 
At $\tau=0$ (and confining ourselves to the winding number zero sector)
we expand
\be\label{ex} X(\s, 0)=x_0+\sum_{m\ne 0}|m|^{-\shalf}(a_me^{-im\s}+\tilde a_me^{im\s})\ee
where we have written (as in \cite{clny}) for later convenience
\be \a_m=-i\sqrt ma_m,~~\a_{-m}=-i\sqrt m a_{-m}, {\rm etc.}\ee
Define also 
$\hat x_m=a_m+\tilde a_m^{\dagger},~~\hat{\ov x}_m=a_m^{\dagger}+\tilde a_m,~m>0$. The eigenstate of these operators which is also an eigenstate of total momentum with eigenvalue zero is,
\be |x,\bar x>=\prod_{m=1}^{\infty}e^{-\shalf\ov x_mx_m-a^{\dag}_m\tilde a_m+a^{\dag}_mx_m+\ov x_m\tilde a^{\dag}_m}|0>\ee
where
$a_m|0>=\tilde a_m |0>=0~m>0,~~\hat p|0>=0$. Note that the first term in
the exponential is simply the bulk bosonic action evaluated with the solution that is regular in the upper half plane (or interior of the disc) 
\be\label{ex2} X(\s,\tau )=x_0+\sum_{m> 0}m^{-\shalf}(x_me^{-im\s}+\ov x_me^{im\s})e^{-m\tau}.\ee

This state is normalized and satisfies the completeness relation,
\be \int [dx][d\ov x]|x,\ov x><x,\ov x|=1,~~[dx]=\prod_{m>1}dx_m.\ee
The boundary state is  then written as 
\be\label{bosebound} |\Psi,b >=\int [dx][d\ov x]e^{-S(x,\ov x)}|x,\ov x>\ee
where $S$ is a boundary action. When the latter is zero we have a state
with N boundary conditions i.e. 
\be |\Psi,b>_0=\int [dx][d\ov x]|x,\ov x >=\prod_{m>0}e^{a_m^{\dag}\tilde a^{\dag}_m}|0>.\ee
For the fermion in the NS sector we have (at $\tau =0$) the expansions,
\be\psi (\s,0)=\sum_{r}\psi_re^{ir\s},~~\tilde\psi (\s,0)=\sum_{r}\psi_re^{ir\s},\ee
where $r\e{\cal Z}-\shalf$. The Majorana conditions give $\psi_{-r}=-\psi_r^{\dag},~\tilde\psi_{-r}=\tilde\psi_r^{\dag}$. The fermionic position eigenstate \cite{clny} is\footnote{We've redefined the $\t$ coordinate in \cite{clny} in order to be able to write the amplitude as a classical action.}
\be\label{fermistate} |\t ,\ov\t;\pm >=\prod_{r>0} exp\{-i\ov\t_r\t_r\pm i\psi_r^{\dag}\tilde\psi_r^{\dag}+i\sqrt 2\psi_r^{\dag}\t_r\mp \sqrt 2 \ov\t_r\tilde\psi_r^{\dag}\}|0>\ee
with $\psi_r|0>=\tilde\psi_r|0>=0$. This state satisfies the boundary conditions
\be (\sqrt 2\ov\t_r-\psi_r^{\dag}\mp i\tilde\psi_r)|\t,\ov\t;\pm >=0,~(i\sqrt 2\t_r-\psi_r\pm i\tilde\psi_r^{\dag})|\t,\ov\t;\pm >=0.\ee
The analog of the bosonic boundary state (\ref{bosebound}) is
\be\label{fermibound} |\Psi,\t >= \int \prod_rd\ov\t_rd\t_r e^{-S(\t )}|\t ,\ov\t >\ee

Now let us put in the tachyon boundary term \cite{hkm}
$$S_T={1\over 8\pi}\int_{\tau =0} d\s (T^2+\t^{\mu}\pa_{\mu}T\pa_{\s}^{-1}\t^{\nu}\pa_{\nu}T )$$ where 
\be\label{theta}\t= \t(\s,\tau )=\sum_{r>0}(\t_re^{ir\s}+\ov\t_re^{-ir\s})e^{-r\tau}\ee
is defined in terms of the boundary coordinates $\t_r$ introduced above and is regular in the upper half $w~(=\s+i\tau)$ plane (or in the interior of the disc $|z|<1$ in terms of the coordinate $z=e^{-iw}$ ).
If we use this expansion for $\t$ we get for the  classical bulk fermionic action
\be\label{fermiaction}S_{\psi}={1\over 2\pi}\int d\s d\tau (\t\pa_w\t+\t\pa_{\ov w}\t)=i\sum_{r>0}\ov\t_r\t_r,\ee
as in $\cite{kl}$. This is the first term in the exponential in the definition of the boundary state (\ref{fermistate}) and it was in order to get this agreement with the bulk action that we redefined $\t$ from that given in \cite{clny}.

Let us now introduce a linear tachyon profile as in \cite{kmm2} and for simplicity take it to be along one coordinate direction.
So \be T=yX\ee
Using the expansions (\ref{ex2}) and (\ref{theta}) we then get
\be\label{tachaction} S_T={u\over 4}(x_0^2+2\sum_{m>0}m^{-1}x_m\ov x_m)+\sum_{r>0}{u\over r}\ov \t_r\t_r\ee
where $u=y^2$.
With this tachyon profile the boundary state (the product of (\ref{bosebound}) and (\ref{fermibound}))  can be easily evaluated since the  integrals are Gaussian. We find\footnote{Ignoring a $u$ independent infinite product of 2's and $\pi$ 's}
\be\label{bound} |B,u>_{NSNS}=\prod_{m=1}^{\infty}\prod_{r=\shalf}^{\infty}
{(1+{u\over r})\over (1+{u\over m})}
e^{-(1-{2\over 1+u/m})a_m^{\dag}\tilde a_m^{\dag}-i(1-{2\over 1+u/r})\psi_r^{\dag}\tilde\psi_r^{\dag}}|0>\int_{-\pi R}^{\pi R} dx_0e^{-{u\over 4}x_0^2}|x_0>\ee
Clearly this boundary state is such that  it satisfies 
Neumann boundary conditions (appropriate to the $\p$ directions) for $u=0$  and  
 Dirichlet conditions (appropriate to the 
$\perp$ directions)   as $u\ra \infty$ (see (\ref{bconditions})). Thus the state (\ref{gsostate}) or  the first term of (\ref{D}) may
be written alternatively as products of (\ref{bound}) with $u=0$ in the
$\p$ directions and $u\ra\infty$ in the $\perp$ directions. The overlap
with the one graviton state  (\ref{overlapD}) is now given by (calling
the ratio of the infinite products $F(u)$
\be\label{}C\prod_{\mu\e\p}(2\pi R^{\mu})\prod_{\mu\e\perp}F(u_{\mu})\int_{-\pi R}^{\pi R}e^{-{u_{\mu}\over 4}(x_0^{\mu})^2}\ee
in the limit $u_{\perp}\ra\infty$. C is essentially the $u$ (and hence $p$) independent constant determined earlier.

The ratio of infinite products in the above  $F(u)$ has been evaluated (after regularization) in \cite{kmm2, kl, ttu} and takes the value
$\sqrt{2\pi}4^uu\Gamma (u)^2\over \Gamma (2u)$.  As  shown in these papers the correct tension ratio  is obtained from this formula. What we have demonstrated above is that this is a consequence of the fact that what is evaluated there is essentially the normalization constant $g_p$ of the boundary state.

\sect{Acknowledgments} 
I wish to thank Joe Polchinski for a discussion. This work is partially supported by
the Department of Energy contract No. DE-FG02-91-ER-40672.

%%%%%%%%%%%%%%%%%%%%%%%%%%%%%%%%%%%%%%%%%%

%%%%%%%%%%%%%%%%%%%%%%%%%%%%%%%%%%%%%%%%%%%%%%%%
\end{document}